\documentclass[traditabstract]{aa}
\usepackage{color}
\usepackage{txfonts,epsfig,graphicx,natbib,twoopt}
\bibpunct{(}{)}{;}{a}{}{,}

\begin{document}
   \title{Discovery of new low-excitation planetary nebulae\
\thanks{Reduced imaging and spectroscopic data are only available at the CDS via anonymous ftp to cdsarc.u-strasbg.fr (130.79.128.5) or via 
http://cdsarc.u-strasbg.fr/viz-bin/qcat?J/A+A/}}

   \author{Chih-Hao Hsia\inst{\ref{inst1}} and Yong Zhang\inst{\ref{inst1}}}

   \titlerunning{Discovery of New Planetary Nebulae}
    \authorrunning{C. -H.~Hsia \& Y.~Zhang}
    \institute{Department of Physics, University of Hong Kong, Pokfulam Road, Hong Kong, China\\
            \email{xiazh@hku.hk}\label{inst1}
              }

   \date{Received xx xxxx 2013 / Accepted 2 January 2014}

\abstract{We report a multi-wavelength study of four new planetary nebula (PN) candidates selected from the INT/WFC Photometric H$\alpha$ 
Survey of the Northern Galactic Plane (IPHAS) and Deep Sky Hunter (DSH) catalogues. We present mid-resolution optical spectra 
of these PNs. The PN status of our sample was confirmed by optical narrow-band images and mid-resolution spectra. Based on the locations 
of these objects in the log (H$\alpha$/[\ion{N}{II}]) versus log (H$\alpha$/[\ion{S}{II}]) diagnostic diagram, we conclude that these 
sources are evolved low-excitation PNs. The optical and infrared appearances of these newly discovered PNs are discussed. 
Three of the new nebulae studied here are detected in infrared and have low infrared-to-radio flux ratios, probably suggesting that 
they are evolved. Furthermore, we derive the dynamical ages and distances of these nebulae and study the spectral energy distribution 
for one of them with extensive infrared archival data. 
}

    \maketitle

\section{Introduction}

Planetary nebulae (PNs) are the evolutionary end products of low- and intermediate-mass (1 $\sim$ 8 M$_\odot$) stars. The
number of PNs detected in the Milky Way is about 2700, although the total population of PN is believed to be about one order of magnitude
higher. The list of detected PNs includes $\sim$1500 in the Galactic PN Catalogue of  \citet{Kohoutek01} and $\sim$1200
PNs listed in the Macquarie-AAO-Strasbourg H$\alpha$ Catalogue of Galactic PNs (MASH; Parker et al. 2006; Miszalski et al. 2008). The
discrepancy between the detected and expected numbers of PNs in the Galaxy is believed to be primarily due to dust extinction, as most of
the PNs are in the Galactic plane. Many evolved PNs have low surface brightness and can be easily missed in the optical surveys.

In this paper, we present a study of four highly obscured PN candidates that are located within both regions of the INT/WFC Photometric 
H$\alpha$ Survey of the Northern Galactic Plane (IPHAS; Drew et al. 2005)/Deep Sky Hunter (DSH) and the {\it Spitzer} Galactic Legacy 
Infrared Mid-Plane Survey Extraordinaire (GLIMPSE)/{\it Wide-field Infrared Survey Explorer} (WISE) infrared surveys. The sample is 
selected from a catalogue of 781 new PN candidates observed by IPHAS \citep{Viironen09a} and a list of 36 un-catalogued Galactic 
nebulae observed from {\it Digitized Sky Survey} (DSS) and {\it Two Micron All Sky Survey} (2MASS) surveys \citep{Kronberger06}. 
Almost all of these objects are located close to the Galactic plane and are very low-brightness nebulae, indicating that they may have 
high dynamic ages or/and suffer larger interstellar extinction. 

One of the scientific goals of the IPHAS is searching for PNs with low surface brightness or faint extended nebulae 
\citep{Mampaso06, Sabin10}. The IPHAS has mapped all northern Galactic plane coverage to $\vert$b$\vert\leqslant$5$^\circ$
by three filters using the Wide Field Camera (WFC) on the 2.5 m Isaac Newton Telescope (INT) at the Observatorio del Roque de los
Muchachos (La Palma, Spain) \citep{Drew05}. The spatial resolution of each image is 0.$\arcsec$33 and the total area of the survey
covers  $\sim$1800 deg$^{2}$. Two broadband Sloan filters (r$\arcmin$ and i$\arcmin$) and one narrow-band H$\alpha$ filter 
($\lambda_{c}$ = 6568 \AA, $\Delta\lambda$ = 95 \AA) were used with corresponding exposure times of 30, 10, and 120 s, respectively. The 
measured magnitude range for point sources in the survey is $r\arcmin\sim$ 13 to 22 mag. The survey provides us an opportunity to 
obtain better estimates of Galactic PN population, the PN distribution density, the PN formation rate, and the chemical abundance of the 
Galaxy. Until now, a number of new PNs \citep{Viironen09b, Sabin10, Sabin11} have been discovered based on results from this survey.

The DSH project is a group involving many amateur astronomers with the goal of discovering new nebulae by searching
optical and near-infrared archival data \citep{Kronberger06, Jacoby10}.

Although the ionized structures of PNs are well determined by the optical observations, many faint evolved Galactic PNs suffer from
high extinction, making them difficult to detect in the optical wavelengths. Unlike optical observations, infrared observations are
hardly affected by dust extinction and therefore can much more effectively detect PNs. The recent infrared observations of
{\it Spitzer} GLIMPSE \citep{Benjamin03, Churchwell09} and the WISE \citep{Wright10} have made possible new ways to study 
Galactic PNs (e.g. Cohen et al. 2005). These results show that the optical appearances of PNs and their infrared counterparts may differ, 
and the infrared images can also help us to understand the distributions of dust components.

The main aim of this present work is to identify the nature of four new IPHAS/DSH PNs by the narrow-band imaging and
follow-up spectroscopic observations. The observations and data reductions are described in Section \ref{s2}. In Section \ref{s3}, we present 
the results of imaging and spectroscopy in the visible and mid-infrared for these objects. 
The log (H$\alpha$/[\ion{N}{II}]) versus log (H$\alpha$/[\ion{S}{II}]) diagnostic diagram is discussed in Section \ref{s4}. In Section 
\ref{s5}, we present the study of spectral energy distribution for one of the new PNs. The dynamical ages and distances of these nebulae are 
discussed in Section \ref{s6}. Finally, the conclusions are given in Section \ref{s7}.

\section{Observations and data reduction}\label{s2}
    
Our sample consists of four objects (DSH J1941.3+2430, IPHASX J194226.1+214522, IPHASX J195248.8+255359, and IPHASX J232713.1+650923)
selected from the DSH and IPHAS surveys as PN candidates. Then, they are abbreviated as PN1, PN2, PN3, and PN4, respectively, in the
following sections. 

\begin{table*}
\caption{Summary of imaging observations.}
\label{tab1}
\centering
\begin{tabular}{cccccc}
\hline\hline
Object & Abbreviation & Observation Date & Filter Name & Seeing & Exposures \\
       &              &                  &             & (arcsec) & (s) \\
\hline
DSH J1941.3+2430 & PN1 & 2010 Aug 15 & [O III] & 2.3 & 900$\times$2  \\
                       &      &             & H$\alpha$+[N II] & 2.1 & 900$\times$3  \\
IPHASX J194226.1+214522 & PN2 & 2010 Aug 14 & [O III] & 2.1 & 1800$\times$2  \\
                       &      &             & H$\alpha$+[N II] & 2.0 & 900$\times$3  \\
IPHASX J195248.8+255359 & PN3 & 2010 Aug 16 & [O III] & 2.0 & 900$\times$2  \\
                       &      &             & H$\alpha$+[N II] & 1.8 & 1800$\times$2  \\
IPHASX J232713.1+650923 & PN4 & 2010 Aug 16 & [O III] & 1.9 & 900$\times$2  \\
                       &      &             & H$\alpha$+[N II] & 2.0 & 900$\times$2  \\
\hline
\end{tabular}
\end{table*}

\begin{table*}
\caption{Summary of spectroscopic observations.}
\label{tab2}
\centering
\begin{tabular}{cccccccc}
\hline\hline
Object & \multicolumn{2}{c}{Coordinate (J 2000.0)} & Observation Date & Wavelength & Dispersion & Width of Slit & Integration Time \\
\cline{2-3}
       &   R.A. & Dec. &                  &   (\AA)    & ($\AA$ pix$^{-1}$) & (arcsec) & (s) \\
\hline
PN1 & 19 41 19.1 & +24 30 53.0 & 2011 Sep 24 & 4750 - 7450 & 2 & 3.6 & 3200$\times$2, 3600  \\
PN2 & 19 42 26.1 & +21 45 22.4 & 2011 Sep 25 & 4750 - 7450 & 2 & 3.6 & 3200$\times$2  \\
PN3 & 19 52 48.8 & +25 53 59.2 & 2011 Sep 26 & 4750 - 7450 & 2 & 3.6 & 2400$\times$3  \\
PN4 & 23 27 12.7 & +65 09 18.7 & 2011 Sep 26 & 4750 - 7450 & 2 & 3.6 & 2400$\times$3  \\
\hline
\end{tabular}
\end{table*}

\subsection{BFOSC narrow-band imaging}

Imaging observations were performed at the {\it 2.16 m Telescope} on the Xing-Long station of the National Astronomy Observatories of
China (NAOC) on the nights of 2010 August 14-16. The BAO Faint Object Spectrograph and Camera (BFOSC) with a 1242 $\times$ 1152 CCD was
used. The field of view (FOV) of the detector is 9.$\arcmin$3 $\times$ 8.$\arcmin$6 with a pixel scale of 0.$\arcsec$45. Our sample
objects were imaged with two narrow-band filters: [\ion{O}{III}] ($\lambda_{c}$ = 5009 \AA, $\Delta\lambda$ = 120 \AA) and
H$\alpha$+[\ion{N}{II}] ($\lambda_{c}$ = 6562 \AA, $\Delta\lambda$ = 140 \AA). The exposure times for these objects ranged from 1800 to 
3600 s. The seeing conditions during the observing run varied between 1.$\arcsec$8 and 2.$\arcsec$3. The data were processed through the 
IRAF software package calibration. Standard bias subtraction and flat-field correction were performed. Data were taken in two-step dithered 
positions to enhance spatial sampling and cosmic rays removal. The journal of observations is summarised in Table \ref{tab1}. 

\subsection{OMR optical spectroscopy}

Mid-resolution spectra of these sources were also obtained by the {\it 2.16 m telescope} of NAOC. An Optomechanics Research Inc. (OMR)
spectrograph and a PI 1340 $\times$ 400 CCD were used, which result in a spectral dispersion of $\sim$ 2.0 \AA~pixel$^{-1}$. The 
spectral coverage of the observations is from 4750 to 7450 $\AA$ at a resolution of 11 $\AA$ FWHM. The seeing conditions varied 
from 2.$\arcsec$0 and 2.$\arcsec$2 during the observing run. A slit width of 3.$\arcsec$6 was set through the main nebulae off the  
geometric centres to avoid possible contamination from the central stars. The length of long slit is 4$\arcmin$ and was placed along 
the north-south direction. Due to the instrumental limitation, the long slit was not rotated to the parallactic angle, tending to lead 
to flux losses caused by atmospheric dispersion \citep{Filippenko82}. However, the effect is negligible, since the slit width is 
large for our observations, and the objects were observed when they were close to the meridian. The exposure times 
ranged from 6400 to 10000 s. Exposures of He-Ar arcs were obtained right before and after each spectrum of standard star and used for 
the wavelength calibration.

Data were reduced following standard procedures in the NOAO IRAF V2.14 software package. The CCD reductions included bias and flat-field
correction, background subtraction, and cosmic-ray removal. Flux calibration was derived from observations of at least three of the KPNO
standard stars per night. The standard stars HR 7059, HR 7596, and HR 9087 were used for flux calibration.
The atmospheric extinction was corrected by the mean extinction coefficients measured for Xing-Long station,
where the {\it 2.16 m Telescope} is located. A final spectrum for each object was produced using the co-added separate exposures to improve
the signal-to-noise ratio and a summary of spectroscopic observations is given in Table \ref{tab2}.

All line fluxes of these objects were measured using Gaussian line profile fitting. The uncertainties in fluxes are obtained from the noise 
level in the continuum. If we take into account the uncertainties of absolute flux calibrations and line flux measurements, the flux 
errors are estimated to be about 15$\%$ - 30$\%$.

\subsection{Spitzer GLIMPSE and MIPSGAL data}\label{s2.3}

Infrared images surrounding the sample objects were extracted from  the {\it Spitzer} GLIMPSE I survey. The GLIMPSE I survey mapped most
of the inner Galactic plane at mid-infrared 3.6, 4.5, 5.8, and 8.0 $\mu$m using the Infrared Array Camera (IRAC; Fazio et al. 2004), which
covers a Galactic region of $\vert$l$\vert$ = 10$^\circ$ - 65$^\circ$ and $\vert$b$\vert\leqslant$ 1$^\circ$ (total area $\sim$ 220
deg$^{2}$). The angular resolutions of the IRAC camera in these four infrared bands were from 1.$\arcsec$6 to 1.$\arcsec$8 with a field of
view of $\sim$ 5.$\arcmin$2 $\times$ 5.$\arcmin$2. The observations were obtained with two 2 s exposures at each position during the periods
between 2004 March and 2004 November. The basic data processing performed by the Spitzer Science Center (SSC) consists of correction for
instrumental artifacts, determination of  the flux densities and positions of point sources, and final mosaic the images. The GLIMPSE I
data yield various products including mosaic images covering the survey area, a high reliable GLIMPSE Point-Source Catalog (GLMC), and a
more complete GLIMPSE Point-Source Archive (GLMA).

Three out of the four nebulae (PN1, PN2, and PN3) in our sample are found to lay within the GLIMPSE I area. Their integrated fluxes 
in each band were measured using the method described in \citet{Kwok08} and \citet{Zhang12}. Two individual square apertures of 
the same size were used to measure the on-nebula ($F_{n}$) and background ($F_{b}$) fluxes. To reduce the measuremental 
errors, we measured the fluxes using the same apertures with different background positions for each target (the used apertures depend on 
the apparent sizes of these nebulae seen in the images), and then the adopted values were obtained from the average of all measurements. 
In all four IRAC images, the apertures were put in the identical positions in each frame. Then the sum of all the fluxes of sources in the 
point-source archive within each aperture were obtained ($F_{n,p}$ and $F_{b,p}$) and subtracted from $F_{n}$ and $F_{b}$, respectively. 
The total fluxes of these three nebulae were obtained from the difference between ($F_{n} - F_{n,p}$) and ($F_{b} - F_{b,p}$). Due to 
relatively large sizes for our objects ($>$ 16$\arcsec$), the extended source aperture calibrations must be made using the correction 
factors \citep{Reach05} suggested by Jarrett\footnote{http://www.ast.uct.ac.za/$\sim$jarrett/irac/calibration/ext$_{-}$apercorr.html.}. 

The PN1, PN2, and PN3 are also detected by the Multiband Imaging Photometer for {\it Spitzer} (MIPS; Rieke et al. 2004)
in the {\it Spitzer} Legacy program MIPS Inner Galactic Plane Survey\footnote{http://mipsgal.ipac.caltech.edu/} (MIPSGAL). The MIPSGAL
observations were taken from different epochs between 2005 and 2006 and produced 24 and 70 $\mu$m imaging with an area of
278 deg$^{2}$, which is located in the inner Galactic plane ($\vert$l$\vert\leqslant$ 62$^\circ$ and $\vert$b$\vert\leqslant$ 1$^\circ$).
In this study, we only used the available 24 $\mu$m images of the same regions for our sample (PN4 does not lay within the MIPSGAL 
field). The 24 $\mu$m fluxes are determined using the same method as described above. The aperture calibrations of these three objects 
at 24 $\mu$m vary in different source colours and apparent sizes. We have corrected the fluxes using the calibration factors 
suggested by MIPS Instrument Handbook\footnote{http://irsa.ipac.caltech.edu/data/SPITZER/docs/mips/mipsinstrument
handbook/1/}. 

The photometric uncertainties of flux measurements in these bands are estimated to be $\sim$ 14$\%$, 17$\%$, 18$\%$, 
12$\%$, and 9$\%$ for 3.6, 4.5, 5.8, 8.0, and 24 $\mu$m, respectively. These values of the uncertainties are derived from the standard 
deviations of flux determinations associated with our aperture photometry method, which are dominated by the uncertainties due to 
background substraction.     

\subsection{WISE data}\label{s2.4}

The WISE mission has mapped the sky in four bands at 3.4, 4.6, 12, and 22 $\mu$m with resolutions from 6.$\arcsec$1 to 12$\arcsec$. The 
preliminary data release includes the first 105 days of mission from 14 January 2010 to 29 April 2010, which covers about 57$\%$ of all sky. 
All the nebulae are located within the area of preliminary data \footnote{http://wise2.ipac.caltech.edu/docs/release/prelim/}, but 
the infrared counterpart of PN4 is too faint to be detected by WISE.  
All four bands were imaged simultaneously, and the exposure times were 7.7 s in 3.4 and 4.6 $\mu$m and 8.8 s in 12 and 22 $\mu$m.
The data presented here were processed with initial calibrations and reduction algorithms. The primary release data products include
10,464 calibrated, mosaic images, a highly reliable Point-Source Catalog containing over 257 million objects detected on the WISE images,
and an extra moving object supplement that provides the positional information of asteroids, comets, and planetary satellites.

Although the WISE 3.4 and 4.6 $\mu$m bands are similar to {\it IRAC} 3.6 and 4.5 $\mu$m channels, the wavelength range of the 3.4
$\mu$m band is slightly bluer compared to {\it IRAC} 3.6 $\mu$m channel and 4.6 $\mu$m band is slightly redder compared to
{\it IRAC} 4.5 $\mu$m band. The WISE 12 $\mu$m band is more sensitive than the {\it IRAC} 8 $\mu$m channel because the WISE
12 $\mu$m passband is significantly broader than that of the {\it IRAC} 8 $\mu$m band. The aromatic infrared band (AIB) features
at 7.7, 8.6, 11.3, 12.7, and 16.4 $\mu$m are covered in the WISE 12 $\mu$m band, but only the first two features lie within
the {\it IRAC} 8 $\mu$m band. Both filters cover a number of nebular atomic emission lines.
The WISE 22 $\mu$m band is less affected by saturation problems compared to the MIPS 24 $\mu$m band because the saturation for point
sources at 22 $\mu$m band is $\sim$ 12.4 Jy, which is higher than that at MIPS 24 $\mu$m band saturation value of 2 Jy (see
WISE explanatory supplement\footnote{http://wise2.ipac.caltech.edu/docs/release/prelim/expsup/
wise$_{-}$prelrel$_{-}$toc.html}).

We perform aperture photometry for all nebulae in our sample using the {\bf phot} task in the APPHOT package of IRAF software (ver. 2.14).
The field stars close to the objects for each band were removed. Then the local background estimations and total flux measurements for
each source per band were obtained. For each band, the photometry was performed five times with different background positions, 
resulting in a standard deviation of flux. The deviations may be high if the nebulae lay within the fields with non-uniform 
background emissions or they have non-circular shapes (circular apertures were used when carrying out the photometry), and vice versa. 
To estimate the uncertainties in flux, we adopt the standard deviations of all background-subtracted flux measurements. The most 
influenced bands in our WISE observations are 12 and 22 $\mu$m channels, and the characteristic uncertainties of all flux measurements 
are estimated to be $\sim$ 18$\%$, 17$\%$, 18$\%$, and 19$\%$ for 3.4, 4.6, 12, and 22 $\mu$m, respectively.
The {\it Spitzer} and WISE infrared photometric results of these PNs are given in Table \ref{tab3}. 

\subsection{Other infrared data}\label{s2.5}

The archival data from the 2MASS \citep{Skrutskie06}, Infrared Astronomical Satellite (IRAS; Neugebauer et al. 1984), and AKARI Infrared 
Astronomical Mission \citep{Murakami07} were also used. All of them mapped entire sky covering the wavelength range from near-infrared 
to far-infrared. The 2MASS survey scanned all objects in three bands at J (1.25 $\mu$m), H (1.65 $\mu$m), and Ks (2.17 $\mu$m) and obtained 
4,121,439 FITS images with the pixel size of $\sim$ 2.$\arcsec$0. The 2MASS Point Source Catalogue \citep{Cutri03} was produced using these 
images and cataloged 470,992,970 sources. The IRAS survey was the first infrared observatory to detect point sources in four broadbands 
at 12, 25, 60, and 100 $\mu$m with resolutions from 30$\arcsec$ (12 $\mu$m) to 2$\arcmin$ (100 $\mu$m). One of the primary products of 
IRAS observation was the point source catalogue (PSC), which contained $\sim$ 246,000 point sources. The AKARI mission had carried out an
all-sky survey in two mid-infrared bands centred at 9 and 18 $\mu$m and four far-infrared channels centred at 65, 90, 140 and 160 
$\mu$m. The angular resolutions of the images in these bands were from 2.$\arcsec$34 to 44.$\arcsec$2 pixel$^{-1}$. The primary product of 
this survey was the point source catalogue, which contained $\sim$ 0.9 million objects that were clearly resolved by the infrared astronomical 
satellite. These archival infrared measurements of the nebulae are given in Table \ref{tab4}.  

\begin{table*}
\caption{{\it Spitzer} and WISE observations.}
\label{tab3}
\centering
\begin{tabular}{cc@{\extracolsep{0.1in}}cccccccc}
\hline\hline
Objects & \multicolumn{5}{c}{Spitzer Flux(mJy)} & \multicolumn{4}{c}{WISE Flux(mJy)} \\
\cline{2-6}\cline{7-10}
 & 3.6$\mu$m & 4.5$\mu$m & 5.8$\mu$m & 8.0$\mu$m & 24$\mu$m & 3.4$\mu$m & 4.6$\mu$m & 12$\mu$m & 22$\mu$m \\
\hline
PN1 & 3.1$\pm$0.4 & 5.6$\pm$0.8 & 7.3$\pm$1.3 & 17.4$\pm$1.5 & 259.1$\pm$25.3 & 2.1$\pm$0.3 & 4.2$\pm$0.4 & 46.8$\pm$7.2 & 242.2$\pm$52.1 \\
PN2 & 5.6$\pm$0.5 & 7.3$\pm$1.3 & 10.8$\pm$1.9 & 16.8$\pm$1.8 & 132.5$\pm$18.4 & 4.1$\pm$0.7 & 6.0$\pm$1.4 & 24.0$\pm$5.4 & 84.7$\pm$19.6 \\
PN3 & 4.6$\pm$0.9 & 9.8$\pm$1.9 & 20.5$\pm$3.8 & 47.5$\pm$8.3 & 786.1$\pm$20.7 & 4.2$\pm$0.9 & 8.7$\pm$1.5 & 116.8$\pm$19.8 & 585.4$\pm$65.1 \\
PN4 & ... & ... & ... & ... & ... & ... & ... & ... & ... \\
\hline
\end{tabular}
\end{table*}

\begin{table*}
\caption{Other flux measurements.}
\label{tab4}
\centering
\begin{tabular}{lc@{\extracolsep{0.1in}}ccc}
\hline\hline
 & \multicolumn{4}{c}{Flux} \\
\cline{2-5}
Filters & PN1 & PN2 & PN3 & PN4 \\
\hline
IPHAS r$\arcmin$$^a$ (mag) & ... & ... & 17.91$\pm$0.01 & ... \\
2MASS J (mag) & 14.85$\pm$0.05 & 16.20$\pm$0.11 & 15.68$\pm$0.07 & ... \\
2MASS H (mag) & 14.40$\pm$0.06 & 15.00$\pm$0.09 & 14.90$\pm$0.08 & ... \\
2MASS Ks (mag) & 14.15$\pm$0.08 & 14.40$\pm$0.11 & 13.77$\pm$0.11 & ... \\
AKARI F18 $\mu$m$^b$ (Jy) & ... & ... & 0.35$\pm$0.03 & ... \\
AKARI F65 $\mu$m$^b$ (Jy) & ... & ... & 6.81:& ... \\
AKARI F90 $\mu$m$^b$ (Jy) & ... & ... & 4.94$\pm$0.80 & ... \\
AKARI F140 $\mu$m$^b$ (Jy) & ... & ... & 0.4: & ... \\
IRAS F12 $\mu$m$^c$ (Jy) & ... & ... & $<$0.31$\pm$0.04 & ... \\
IRAS F25 $\mu$m$^c$ (Jy) & ... & ... & 0.69$\pm$0.20 & ... \\
IRAS F60 $\mu$m$^c$ (Jy) & ... & ... & 3.89$\pm$0.80 & ... \\
IRAS F100 $\mu$m$^c$ (Jy) & ... & ... & $<$36.87$\pm$2.30 & ... \\
NVSS 1.4GHz (mJy)& 7.40$\pm$0.50 & 5.50$\pm$0.50 & 15.90$\pm$0.60 & 2.50$\pm$0.50 \\
\hline
\end{tabular}
\tablefoot{
\tablefoottext{a}{From \citet{Viironen09a}.}
\tablefoottext{b}{From AKARI database. The colon represents unreliable detection.}
\tablefoottext{c}{From IRAS database. For 12 $\mu$m and 100 $\mu$m measurements, the upper limits of fluxes are given.}
}
\end{table*}
    
\subsection{Radio data}

Most PNs are weak radio continuum sources due to free-free emissions from their ionized components. We made use of the 1.4 GHz radio
fluxes from the National Radio Astronomy Observatory (NRAO) Very Large Array (VLA) Sky Survey (NVSS; Condon \& Kaplan 1998). All four 
PNs can be detected in the NVSS radio survey. The NVSS covers most of the entire sky ($\delta\geqslant$ -40$^\circ$) with a 
sensitivity of $\sim$ 2.5 mJy beam$^{-1}$ at 1.4 GHz, which provides a spatial resolution of 45$\arcsec$, so that all objects in our sample 
can not be spatially resolved by NVSS. 

\section{Results for new planetary nebulae}\label{s3}

Results from the imaging, spectroscopic, and photometric studies in the visible, infrared, and radio of these four new PNs are presented below. 

\begin{figure}
\begin{center}
\includegraphics[width=0.5\textwidth]{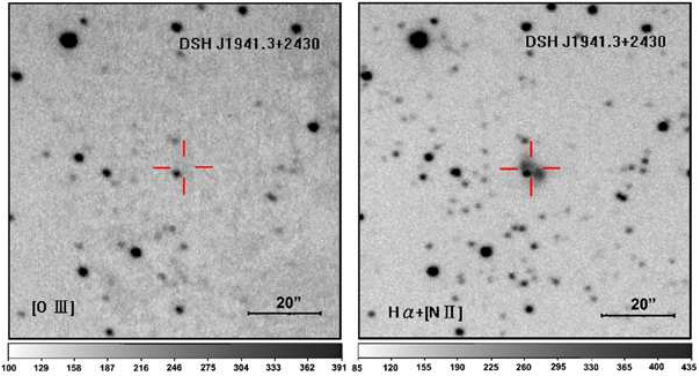}
\end{center}
\caption{Images of DSH J1941.3+2430 (PN1) in [\ion{O}{III}] (left) and H$\alpha$+[\ion{N}{II}] (right) displayed on a linear intensity scale.
North is up and east is to the left. The gray-scale bar is given at the bottom in units of counts per pixel. The position of this PN is
marked with the red central cross.}
\label{pn1i}
\end{figure}

\begin{figure}
\begin{center}
\includegraphics[width=0.5\textwidth]{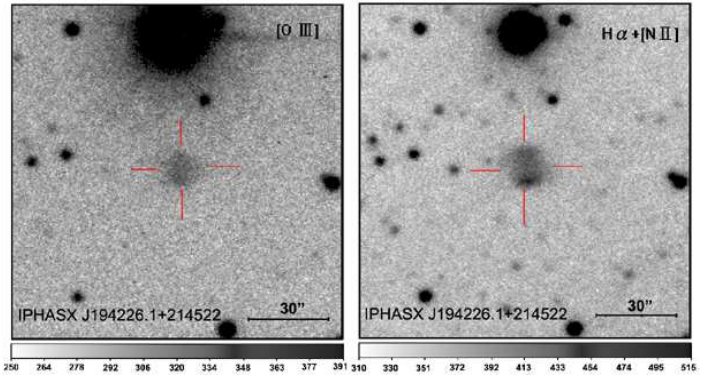}
\end{center}
\caption{Images of IPHASX J194226.1+214522 (PN2) in [\ion{O}{III}] (left) and H$\alpha$+[\ion{N}{II}] (right) displayed on a linear 
intensity scale. North is up and east towards the left. The  gray-scale bar is given at the bottom in units of counts per pixel. 
The position of PN2 is marked with the red central cross.}
\label{pn2i}
\end{figure}

\begin{figure}
\begin{center}
\includegraphics[width=0.5\textwidth]{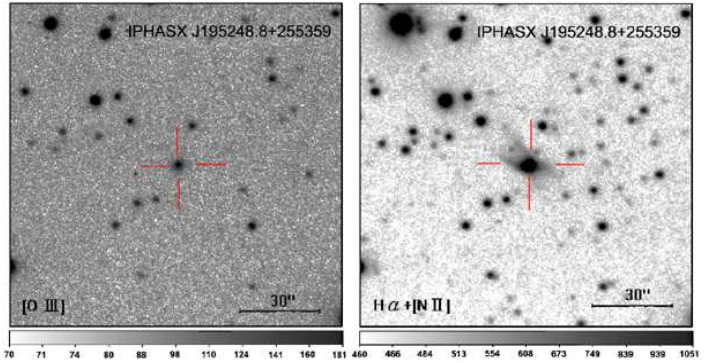}
\end{center}
\caption{The [\ion{O}{III}] (left) and H$\alpha$+[\ion{N}{II}] (right) images of IPHASX J195248.8+255359 (PN3) displayed on a logarithmic
intensity scale to show the outer lobes of the nebula. North is up and east is to the left. The gray-scale bar is given at the bottom in
units of counts per pixel. The position of PN3 is marked with the red central cross.}
\label{pn3i}
\end{figure}

\begin{figure}
\begin{center}
\includegraphics[width=0.5\textwidth]{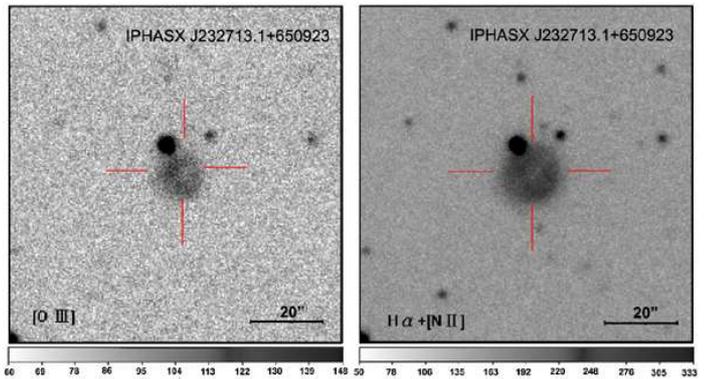}
\end{center}
\caption{Images of IPHASX J232713.1+650923 (PN4) in [\ion{O}{III}] (left) and H$\alpha$+[\ion{N}{II}] (right) displayed on a linear intensity
scale. North is up and east is to the left. The gray scale bar is given at the bottom in units of counts per pixel. The position of PN4 is 
marked with the red central cross.}
\label{pn4i}
\end{figure}

\subsection{Morphology}

The locations of all nebulae in our sample are close to the Galactic plane, which is also where diffuse \ion{H}{II} regions, supernova
remnants (SNRs), star-forming complexes, and other nebulous objects are located. To confirm the PN nature of the candidates,
it would be useful for obtaining narrow-band optical images of the surrounding environment of the objects.

In Figures~\ref{pn1i}-\ref{pn4i}, we show the narrow-band ([\ion{O}{III}] and H$\alpha$+[\ion{N}{II}]) images of the sample objects. All are
well resolved and show distinct morphologies. Their central stars are neither detected nor are very faint in the visible. Two
nebulae  (PN2 and PN3) show bipolar morphology, one (PN4) with round shape and another one (PN1) with irregular structures in 
the H$\alpha$+[\ion{N}{II}] images. The nebulae PN2 and PN4 all show similar appearances in both [\ion{O}{III}] and 
H$\alpha$+[\ion{N}{II}] images, whereas PN3 only reveals the brighter central core in the [\ion{O}{III}] image.

The nebula PN1 was first detected as a PN candidate by \citet{Kronberger06} based on DSS optical imaging and 2MASS observations. Our
H$\alpha$+[\ion{N}{II}] image shows a pair of irregular nebular knots extending $\sim$ 10$\arcsec$ towards northeast - southwest direction
in Figure~\ref{pn1i}. No [\ion{O}{III}] emission is detected associated with this source.

PN2 shows a bright, filled, central waist with two faint bipolar outer lobes  (Figure~\ref{pn2i}). The major axis of the lobes is oriented
at position angle (PA) = 93$^\circ$ and has a length of 16$\arcsec\times$12$\arcsec$ in [\ion{O}{III}] and 23$\arcsec\times$16$\arcsec$ 
in H$\alpha$+[\ion{N}{II}].

PN3 has a typical morphology of a bipolar PN. It has a bright unresolved core and a pair of faint extended bipolar lobes
(Figure~\ref{pn3i}). The extended bipolar lobes are only visible in the H$\alpha$+[\ion{N}{II}] image but not in the [\ion{O}{III}] image.
The major axis of the bipolar lobes has a position angle of 49$^\circ$ and the nebula has projected sizes of 6$\arcsec$.5 in [\ion{O}{III}]
and 26$\arcsec\times$9$\arcsec$.5 in H$\alpha$+[\ion{N}{II}], respectively.

Both the [\ion{O}{III}] and H$\alpha$+[\ion{N}{II}] images of PN4 reveal a round appearance (Figure~\ref{pn4i}). The H$\alpha$+[\ion{N}{II}]
appearance is clearly more extended than its [\ion{O}{III}] counterpart. The nebula has a typical round shape with a central cavity and has
a diameter of 18$\arcsec$ in H$\alpha$+[\ion{N}{II}] and 15$\arcsec$ in [\ion{O}{III}].

\begin{figure*}
\begin{center}
\includegraphics[width=0.9\textwidth]{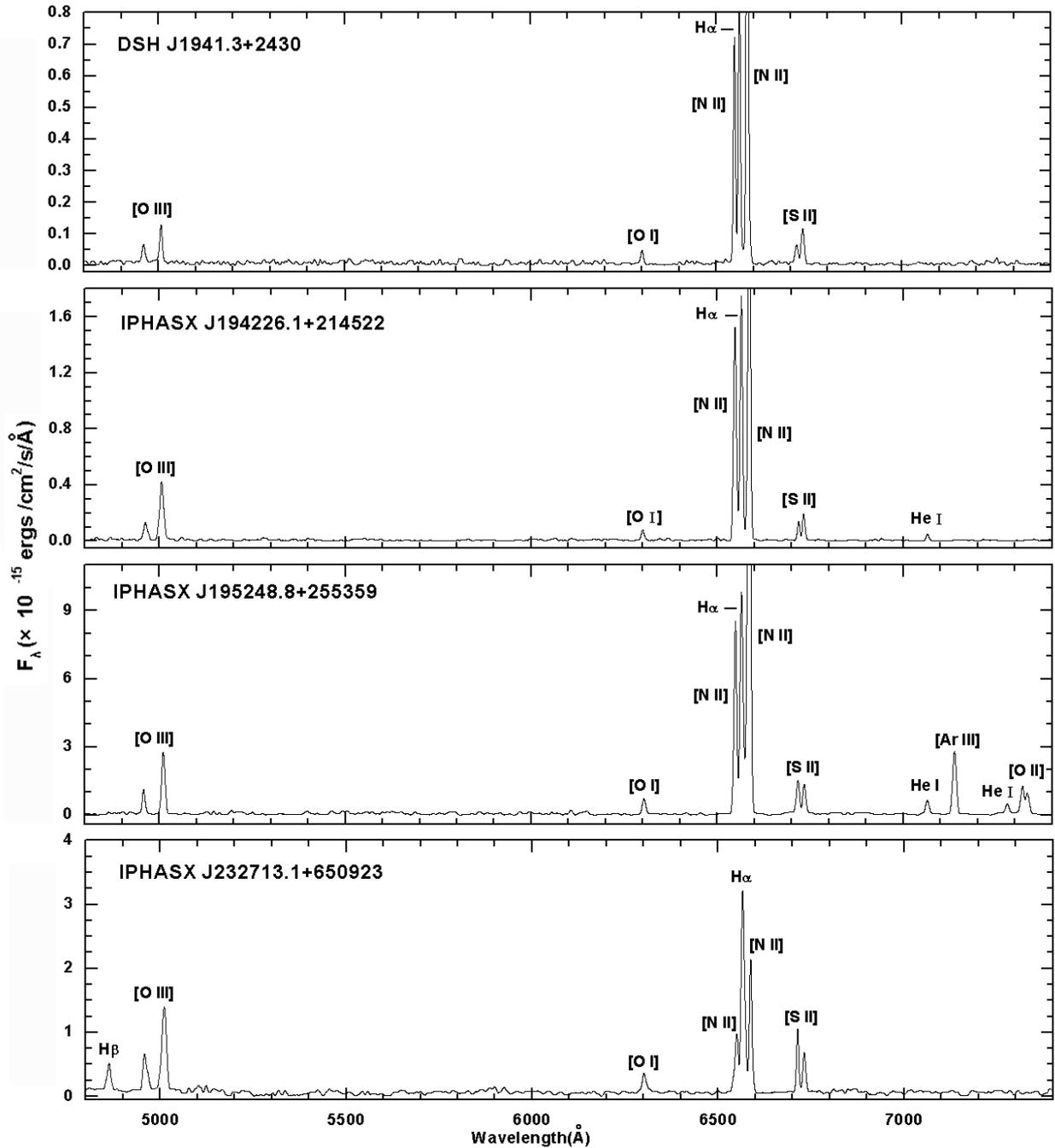}
\end{center}
\caption{Optical spectra of the four sample PNs in the wavelength range of  4800 $\AA$ to 7400 $\AA$. The prominent emission features
are marked.}
\label{spectra}
\end{figure*}

\subsection{Spectral properties}

Figure~\ref{spectra} shows the OMR spectra of the four sample objects. All the nebulae show typical emission lines ([\ion{O}{III}] 
$\lambda\lambda$5007, 4959, H$\alpha$, and [\ion{N}{II}] $\lambda\lambda$6548, 6584) of general PNs, confirming the PN nature
of these objects.  The measured emission line fluxes in the spectra are listed in Table \ref{tab5}. The central wavelengths of the
emissions and line identifications are listed in Column 1 and 2. Columns 3 - 6 give the observed fluxes (normalized to F(H$\alpha$) 
= 100) measured using the Gaussian fitting routine for these objects. Column 7 gives the dereddened flux for PN4, the only PN for 
which we could measure the H$\beta$ line. 
The raw integrated H$\alpha$ fluxes measured from main nebulae for PN1, PN2, PN3, and PN4 are 6.23 $\times$ 10$^{-15}$, 1.46 $\times$
10$^{-14}$, 9.97 $\times$ 10$^{-14}$, and 3.51 $\times$ 10$^{-14}$ erg cm$^{-2}$ s$^{-1}$, respectively. By comparing the observed
H$\beta$/H$\alpha$ intensity ratio with the theoretical value at $T_{e}$ = 10$^{4}$ K and $n_{e}$ = 10$^{4}$ cm$^{-3}$ 
\citep{Hummer87} and by using the reddening law of \citet{Howarth83} for R$_{V}$ = 3.0, we can derive the extinction coefficient 
$c_{H\beta}$ = 1.63$\pm$1.01 for PN4. Although H$\beta$ lines are not detected in PNs 1-3, a 3$\sigma$ lower limit of extinction 
value can be estimated: $F$(H$\alpha$)/$F$(H$\beta$) $>$ 10.6, giving E(B-V)$>$ 1.39. Therefore, all the PNs are highly obscured. 
The line ratios of [\ion{O}{III}] $\lambda$5007/$\lambda$4959 of all the objects are in the range of 2.7 to 3.2, and [\ion{N}{II}]
$\lambda$6584/$\lambda$6548 in the range of 2.8 to 3.3, which in good agreement with the theoretical predictions (e.g. Storey \& Zeippen 
2000). The emissions of \ion{He}{I}
at $\lambda\lambda$7065, 7281 $\AA$ and the intermediate excitation line [\ion{Ar}{III}] at $\lambda$7135 $\AA$ can only be seen in PN3.
Using the [\ion{S}{II}] $\lambda$6731/$\lambda$6717 line ratios and assuming an electron temperature $T_{e}$ = 10000 K, we derive the
electron densities log $n_{e}$ = 3.73$^{+0.54}_{-0.30}$, 3.43$^{+0.40}_{-0.32}$, 2.60$^{+0.32}_{-0.56}$, and 2.48$^{+0.18}_{-0.40}$ 
cm$^{-3}$ for PN1, PN2, PN3, and PN4, respectively.

\begin{table*}
\caption{Measured ($F_\lambda$) and dereddened ($F_\lambda^\prime$) emission line fluxes.}
\label{tab5}
\centering
\begin{tabular}{llccccc}
\hline\hline
 & & \multicolumn{5}{c}{Flux} \\
\cline{3-7}
Wavelength(\AA) & Identification & PN1 & PN2 & PN3 & \multicolumn{2}{c}{PN4} \\
\cline{6-7}
 & & $F_\lambda$ & $F_\lambda$ & $F_\lambda$ & $F_\lambda$ & $F_\lambda^\prime$ \\
\hline
4861 & H$\beta$ & ... & ... & ... & 12.2 (19) & 40.9 (19) \\
4959 & [O III] & 8.4 (20) & 10.9 (23) & 9.8 (19) & 18.7 (18) & 57.3 (19) \\
5007 & [O III] & 23.1 (19) & 34.1 (20) & 31.1 (19) & 54.9 (18) & 160.8 (19) \\
6300 & [O I] & 6.1 (20) & 4.4 (21) & 7.8 (21) & 10.5 (20) & 12.2 (20) \\
6548 & [N II] & 71.8 (22) & 75.4 (19) & 81.2 (21) & 25.7 (26) & 25.9 (26) \\
6563 & H$\alpha$ & 100.0 (20) & 100.0 (23) & 100.0 (19) & 100.0 (19) & 100.0 (19) \\
6584 & [N II] & 210.4 (22) & 209.5 (20) & 262.9 (18) & 76.4 (21) & 75.5 (22) \\
6717 & [S II] & 10.1 (21) & 7.3 (22) & 14.8 (23) & 21.6 (18) & 19.9 (18) \\
6731 & [S II] & 17.8 (22) & 10.9 (24) & 13.4 (22) & 18.3 (18) & 16.8 (18) \\
7065 & He I & ... & 2.8 (20) & 7.2 (21) & ... & ... \\
7135 & [Ar III] & ... & ... & 35.2 (20) & ... & ... \\
7281 & He I & ... & ... & 5.6 (22) & ... & ... \\
7319 & [O II] & ... & ... & 12.6 (19) & ... & ... \\
7330 & [O II] & ... & ... & 11.2 (22) & ... & ... \\
\hline
& & \multicolumn{5}{c}{Line ratio} \\
\hline
H$\alpha$/[S II] & & 3.58 & 5.49 & 3.55 & ... & 2.72  \\
H$\alpha$/[N II] & & 0.35 & 0.35 & 0.29 & ... & 0.99  \\
\hline
\end{tabular}
\tablefoot{The line fluxes were normalized to F(H$\alpha$)=100 and percentage errors of the flux measurements are given within 
brackets.\\}
\end{table*}

\subsection{Infrared properties}

\begin{figure}
\begin{center}
\includegraphics[width=0.5\textwidth]{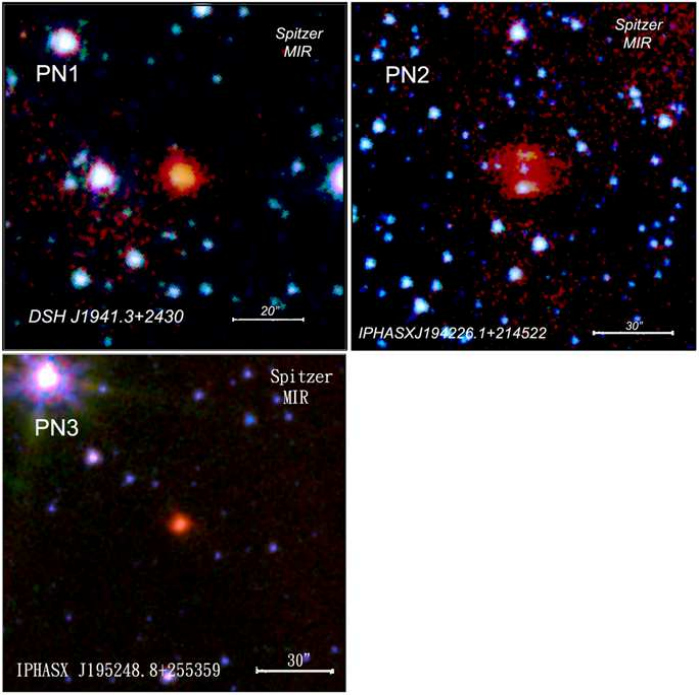}
\end{center}
\caption{Colour composite of the {\it Spitzer IRAC} images of PN1, PN2, and PN3. They are displayed on a logarithmic scale. These infrared
images were made from three {\it IRAC} bands: 3.6 $\mu$m  (shown as blue), 4.5 $\mu$m (green), and 8.0 $\mu$m (red). North is up
and east is to the left. The central star of IPHASX J194226.1+214522 (PN2) can be seen in the {\it IRAC} image.}
\label{irpn}
\end{figure}

To investigate the dust properties of these PNs, we have made use of infrared data from the {\it Spitzer} and WISE data archives.
Three of our objects are in both of the GLIMPSE/MIPSGAL and WISE fields, and all of them have obvious infrared counterparts.
The {\it IRAC} images of these three nebulae are displayed in Figure~\ref{irpn}. The infrared images show that these PNs are clearly much
redder than the field stars, suggesting that they are dusty or reveal infrared emission features, such as AIBs, hydrogen lines, and 
fine-structure lines. We also note that the larger PNs show lower infrared surface brightnesses compared to that with smaller ones. 
The nebula PN1 clearly shows a bipolar/elliptical shape in the infrared while its optical morphology is different. Except 
for the bipolar/elliptical PN1, the infrared and optical morphologies of these objects appear to be similar. The {\it IRAC} images of PN2 
and PN3 both clearly show prominent infrared emissions in the central dust torus/core and a pair of faint lobes, which suggest that most 
dust masses of these PNs are located in the central parts. The bipolar lobes of PN2 and PN3 can be only been seen at 8 $\mu$m.

The {\it MIPS} and WISE detectors have different sensitivities, and the filters have different bandpasses. Both bands have similar 
spectral coverage. We examine the reliabilities of flux calibration and measurement for new PNs in our sample by comparing the MIPS 24 
$\mu$m and WISE 22 $\mu$m integrated fluxes. We obtain the WISE 22$\mu$m/MIPS 24$\mu$m flux ratio of 0.75 $\pm$ 0.14, which is in 
reasonable agreement with the previous results deduced by \citet{Anderson12} (1.01 $\pm$ 0.23).

\subsection{Comparison with mid-infrared and radio flux densities}

\begin{figure*}
\begin{center}
\includegraphics[width=0.9\textwidth]{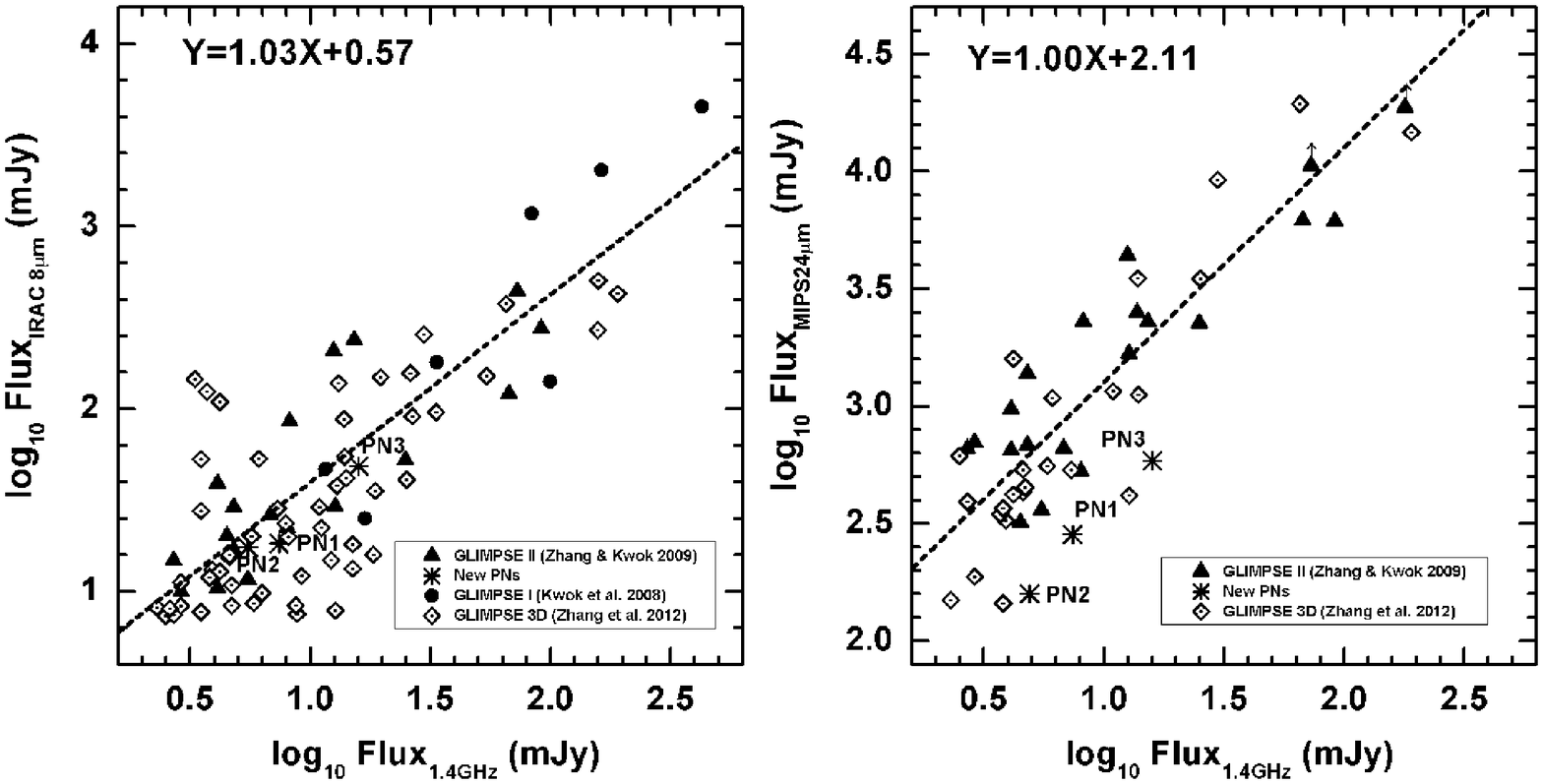}
\end{center}
\caption{
Left panel: {\it IRAC} 8.0 $\mu$m vs. NVSS 1.4 GHz integrated fluxes for 83 PNs. Right panel: MIPS 24 $\mu$m vs. NVSS 1.4 GHz integrated
fluxes for 48 PNs. The dotted line is a linear fit to the data. The fitting parameters are shown on the upper left corner of the figures.
The open diamonds, filled triangles and circles are from the GLIMPSE 3D \citep{Zhang12}, GLIMPSE II \citep{Zhang09}, and GLIMPSE I
\citep{Kwok08} survey, respectively. The asterisks denote our newly discovered PNs.}
\label{nvss}
\end{figure*}

The observed mid-infrared emissions from PNs originate in circumstellar dust components heated by ultraviolet and visible photons from
the central stars, whereas the thermal radio continuum originate in the ionized gas nebulae photoionized by Lyman-continuum photons
from the central stars. As the central star evolves, the ratio of Lyman continuum to sub-Lyman continuum photons increases with time. It
is therefore expected that the infrared to radio flux ratio will also change as the PN evolves \citep{Cohen07, Kwok89, Ortiz11}. 
\citet{Cohen07} found that the infrared-to-radio flux ratio progressively declines with PN evolution. However, their subsequent study 
\citep{Cohen11} does not support the previous conclusion. \citet{Cohen11} suspect that the {\it Midcourse Space Experiment} 
(MSX) infrared fluxes they measured were contaminated by other field stars, and then the infrared-to-radio flux ratios were 
overestimated. It would be useful to investigate if there is any difference between the infrared-to-radio flux ratios of these 
new PNs and that of other PNs closed to Galactic plane (GLIMPSE I, II, and 3D PNs).   
A comparison of the correlations between {\it IRAC} 8 $\mu$m versus NVSS 1.4 GHz and MIPS 24 $\mu$m versus NVSS 1.4 GHz integrated fluxes 
of the PNs is given in Figure~\ref{nvss}. The slopes of 1.03 for {\it IRAC} 8 $\mu$m
to NVSS 1.4 GHz plot and 1.00 for MIPS 24 $\mu$m to NVSS 1.4 GHz plot in the logarithmic space are derived. From the left panel of 
Figure~\ref{nvss}, we find no obvious difference in the trend of these three PNs compared to that of the other PNs. 
Nevertheless, compared to the PNs in the 24 $\mu$m MIPS to 1.4 GHz NVSS plot, the infrared-to-radio flux ratios of these new discovered 
PNs are generally lower than that of the GLIMPSE II and GLIMPSE 3D PNs \citep{Zhang09, Zhang12} (the right panel of Figure~\ref{nvss}). 
Unlike the 8 $\mu$m band, the 24 $\mu$m band does not suffer from the AIB contaminations and can better trace the dust emission. 
It may therefore be expected that these new PNs are more evolved than other PNs in the GLIMPSE II and GLIMPSE 3D fields.

\section{The log (H$\alpha$/[\ion{N}{II}]) versus log (H$\alpha$/[\ion{S}{II}]) diagnostic diagram}\label{s4}

\begin{figure}
\begin{center}
\includegraphics[width=0.48\textwidth]{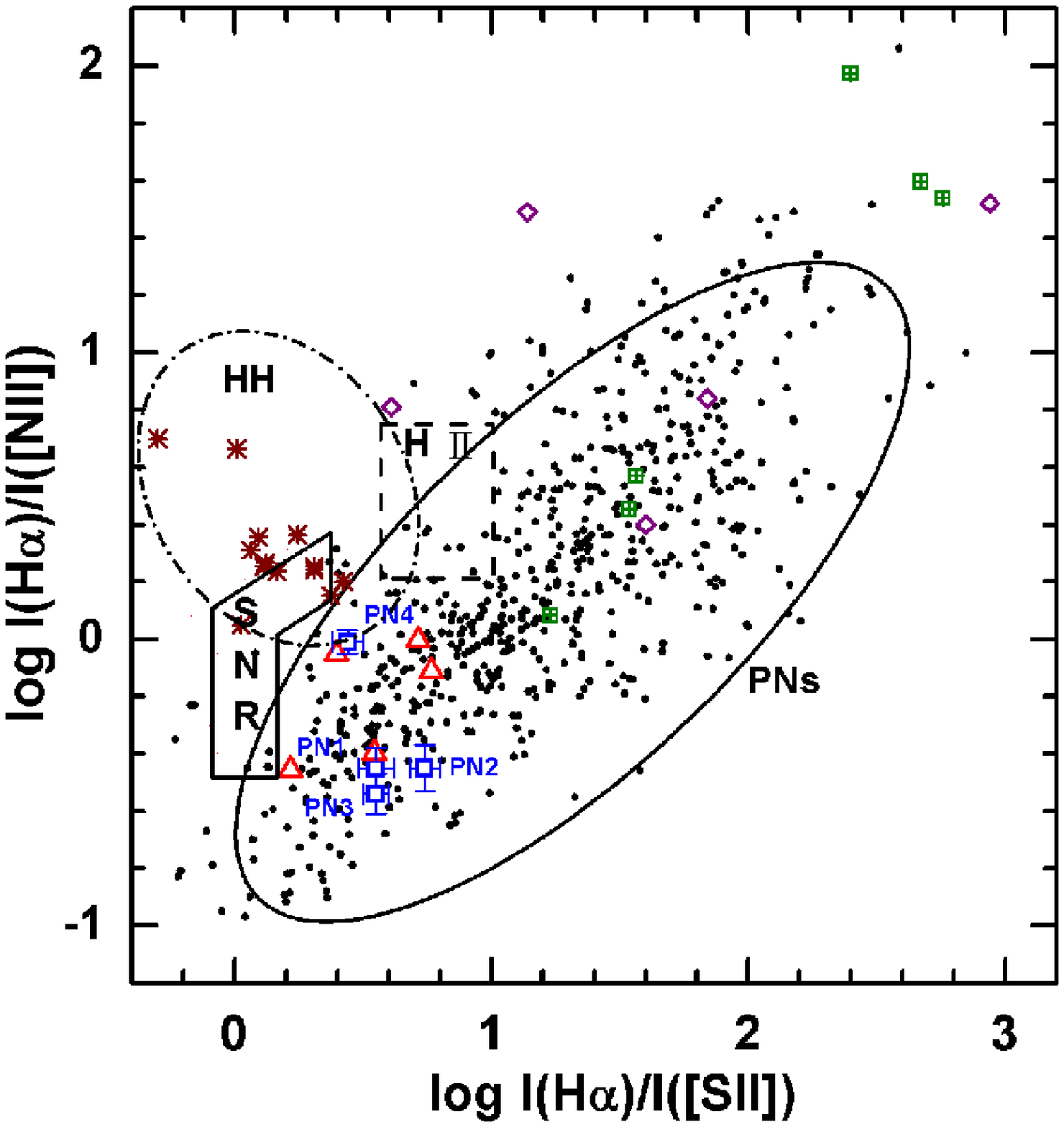}
\end{center}
\caption{Diagnostic diagram for the log (H$\alpha$/[\ion{N}{II}]) intensity ratio versus log (H$\alpha$/[\ion{S}{II}]) ratio showing the
locations of SNRs, \ion{H}{II} regions (Riesgo \& L\'{o}pez 2006), HH objects (Cant\'{o} 1981), and the 90$\%$ probability ellipse of normal
PNs. Also plotted are 551 Galactic PNs (black dots; Acker et al. 1992), HH object sample (brown asterisks; Raga et al. 1996), symbiotic
stars (green squares; Luna \& Costa 2005), IPHAS young PNs (purple diamonds; Viironen et al. 2009b) and five evolved PNs A35, NGC 650-1,
NGC 6583, NGC 7293, and S 176 (red triangles; Sabbadin et al. 1977; Acker et al. 1992) with these new PNs of our sample (blue squares).}
\label{diagnostic}
\end{figure}

The emission-line ratio diagrams are useful probes to the physical conditions of ionized gaseous nebulae. \citet{Sabbadin77} first 
introduced the log(H$\alpha$/[\ion{N}{II}]) vs. log(H$\alpha$/[\ion{S}{II}]) (S2N2) diagnostic diagram to distinguish PNs from \ion{H}{II} 
regions and SNRs. The S2N2 diagnostic diagrams are also useful in differentiating excitation conditions. For example, SNRs are excited by 
the expanding shock waves, and their low H$\alpha$/[\ion{S}{II}] ratios are mainly dominated by the surrounding swept-up ISM. The \ion{H}{II} 
regions are photoionized, extended nebulae with low densities and their locations on the S2N2 diagnostic diagram are mainly influenced 
by their metallicities and excitation conditions (e.g. Viironen et al. 2007). The PNs are commonly nitrogen-rich as the result of CNO-processed 
ejecta from their progenitor stars. From this S2N2 diagnostic diagram, it is clear that the positions of PNs are systematically lower 
than the majority of SNRs, HH objects, and \ion{H}{II} regions, though there are some overlaps in these regions. The PNs also show a wide 
distribution in the H$\alpha$/[\ion{S}{II}] and H$\alpha$/[\ion{N}{II}] ratios depending on their evolutionary stages, metallicities, and 
different observed spatial positions of the nebulae. The ratios are essentially independent of extinction because of marginal difference 
in wavelengths.

To confirm the PN nature of our sample, we placed these objects into a revised version of the log(H$\alpha$/[\ion{N}{II}]) versus
log(H$\alpha$/[\ion{S}{II}]) diagnostic diagram \citep{Riesgo06} (Figure~\ref{diagnostic}).  In the figure, we also plot examples of
Herbig-Haro (HH) objects \citep{Raga96}, symbiotic stars \citep{Luna05}, Galactic PNs \citep{Acker92}, IPHAS young PNs \citep{Viironen09b} 
and five evolved PNs \citep{Sabbadin77, Acker92} for comparison. All the PNs in our sample are located within the 90$\%$ probability
ellipse area of PNs and are well separated from other objects. In addition, most new PNs share a similar position of bottom-left side in
the diagram with other five evolved nebulae, indicating that they are probably low excitation PNs with high dynamic ages. Unlike the
evolved PNs, the young PNs observed by \citet{Viironen09b} are located in top-right side in the diagram.

\section{Spectral energy distribution}\label{s5}

\begin{figure}
\begin{center}
\includegraphics[width=0.5\textwidth]{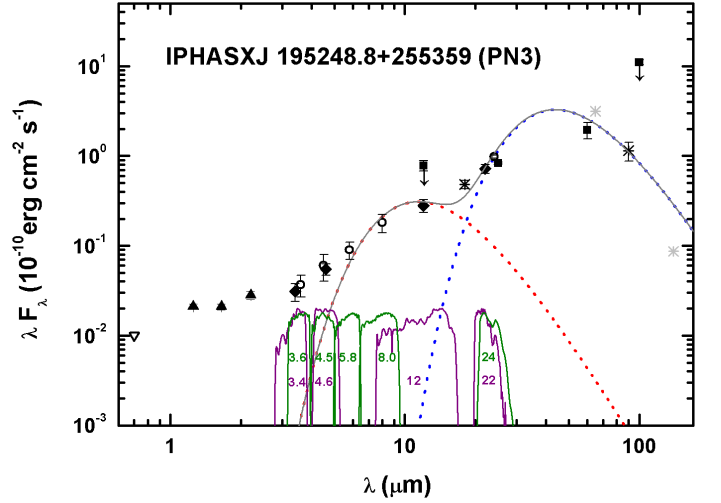}
\end{center}
\caption{The SED of IPHASX J195248.8+255359 (PN3) without extinction correction in the wavelength range form 0.6 to 170 $\mu$m. The
open triangle, filled triangles, open circles, filled diamonds, filled squares, and asterisks are from the r$\arcmin$-band
photometry, 2MASS, {\it Spitzer}/GLIMPSE, WISE, IRAS, and {\it AKARI} survey, respectively. The light asterisks represent the unreliable
{\it AKARI} detections. Note that the fluxes measured from IRAS 12 and 100 $\mu$m are upper limits. The dust component is fitted by
two modified blackbodies (blue and red dashed lines) and total flux of dust component is plotted as a gray solid line. 
The normalized relative spectral response curves for Spitzer 3.6, 4.5, 5.8, 8.0 and 24$\mu$m channels (green lines) and WISE 
3.4, 4.6, 12 and 22$\mu$m bands (purple lines) are also plotted on bottom.}
\label{sed}
\end{figure}

Although PNs are best known by their emission-line spectra, a significant amount of the energy output is in the infrared, emitted 
by the dust component. The contribution of the dust component to the nebular emission can be inferred from the spectral energy distribution 
(SED). In Figure~\ref{sed}, we construct the SED of one of the objects (PN3), which has extensive measurements from infrared archival data. 
In addition to the {\it Spitzer} and WISE photometric measurements, we also used observations from the {\it IRAS} and {\it AKARI} Point 
Source Catalogs \citep{Ishihara10, Helou88} in the infrared regions. Near-infrared magnitudes of J, H, and Ks observed by 2MASS are 
derived from 2MASS All Sky Catalog of Point Source database \citep{Cutri03}. The IPHAS r$\arcmin$ band magnitude of this PN 
is obtained from \citet{Viironen09a}. A summary of the archival data and NVSS measurements is also given in Table \ref{tab4}. Since some of 
the filters used are broad, colour corrections may be needed for some flux density values given in Figure~\ref{sed} and Table \ref{tab4}. 
In the SED fitting, the colour corrections on the calibrated fluxes for WISE bands were performed using the colour correction factors given
by \citet{Wright10}. The colour corrections are dependent on the dust temperature and the intrinsic SED and reflect the deviations of nebular 
spectra from that defined by the photometric system. The colour corrections are generally small except the wide WISE 12 $\mu$m band.

The SED of this PN without extinction correction is shown in Figure~\ref{sed}. A rise in flux toward short wavelengths can be found in 
the SED. This could be due to the contribution from the photospheric and ionized gas components (see Zhang \& Kwok 1991; Hsia et al. 
2010). Due to the lack of ultraviolet observations and the extinction coefficient of this object, the central star and gaseous nebula 
contributions can not be accurately derived. We can see that the observed 
dust emission component is too broad to be fitted by a single dust temperature. Instead, we fit the dust component by two modified 
blackbodies with different temperatures, a warm ($T_{wd}$) and a cold ($T_{cd}$) dust components. The thermal emission from the dust 
component is therefore given by

\begin{equation}
F_\lambda(d)=F_\lambda(wd)+F_\lambda(cd)=\frac{3M_{wd} Q_\lambda B_\lambda(T_{wd})}{4a \rho_{s} D^2} + 
\frac{3M_{cd} Q_\lambda B_\lambda(T_{cd})}{4a \rho_{s} D^2},
\end{equation}

where $F_\lambda(d)$ is the flux density of dust thermal emission continuum, $M_{wd}$ and $M_{cd}$ are the masses of warm and cold dust
components, $\rho_{s}$ is the grain density, $a$ is the radius of grain, which depends on its shape, $Q_\lambda$=Q$_{0}$
($\lambda$/$\lambda_{0}$)$^{-\alpha}$ is the grain emissivity function, and $D$ is the distance to the nebula \citep[eq.~(10.23)]{Kwok07}.
The temperatures of two dust components used in the fittings are 231 K and 60 K. From the SED, we can see that the {\it IRAC} 
3.6/4.5 $\mu$m and WISE 3.4/4.6 $\mu$m bands are mainly dominated by the nebular continuum emission or photospheric contribution. It is 
also clear that the contributions of the WISE 12 $\mu$m and 22 $\mu$m bands are, respectively, from warm and cool dust components. The size 
of a graphite grain ranges from 0.005 $\mu$m $<$ $a$ $<$ 0.25 $\mu$m \citep{Mathis77}. The density of small organic grains ($\rho_{s}$) is 
about 1 g cm$^{-3}$ \citep{Dopita00}. The $\alpha$ value is adopted from 1 to 2 \citep{Stasinska99}. The value of $Q$ (1 $\mu$m) ranges
from 0.04 to 0.8 (which depends on the grain properties; see Figures 4 and 5 of Draine \& Lee (1984)). Assuming $\alpha$ =1.5, 
$\rho_{s}$=1 g cm$^{-3}$, $a$=0.1 $\mu$m, $Q$ (1 $\mu$m)=0.1, and $D$=5.56 kpc (see Section \ref{s6}), we derive the masses of the 
warm and cold dust components to be 1.13 $\times$ 10$^{-7}$ M$_{\sun}$ and 1.87 $\times$ 10$^{-3}$ M$_{\sun}$, respectively. 
Comparison of the dust mass of a typical Galactic PN of $\sim$ 10$^{-4}$ M$_{\sun}$ \citep{Pottasch84}, the dust mass of this nebula is 
significant larger than the usual values, suggesting that this PN is indeed very dusty and its optical counterpart is heavily 
obscured by circumstellar dust. 

\begin{table*}
\caption{Properties of New PNs.}
\label{tab6}
\centering
\begin{tabular}{lcccc}
\hline\hline
Parameters & PN1 & PN2 & PN3 & PN4 \\
\hline
Morphology$^a$ & B or E$^b$ & B & B & R \\
Size [O III] ($\arcsec$) & ... & 16$\times$12 & 6.5 & 15 \\
Size H$\alpha$ ($\arcsec$) & 10 & 23$\times$16 & 26$\times$9.5 & 18 \\
$\theta_{\rm \scriptscriptstyle 8.0}$ ($\arcsec$) & 6.4 & 13 & 5.5 & ... \\
E(B-V) & ... & ... & ... & 1.13$\pm$0.68 \\
$n_e$$^c$ (cm$^{-3}$) & 5400$^{+13400}_{-2700}$ & 2700$^{+4100}_{-1400}$ & 400$^{+430}_{-290}$ & 300$^{+150}_{-180}$ \\
$D_{\rm \scriptscriptstyle MIR}$ (kpc) & 6.24$\pm$3.12 & 4.76$\pm$2.38 & 5.56$\pm$2.78 & ... \\
Age (yr) & 6600$\pm$3300 & 9900$\pm$4950 & 10400$\pm$5300 & ... \\
\hline
\end{tabular}
\tablefoot{
\tablefoottext{a}{B: bipolar; E: elliptical; R: round.}
\tablefoottext{b}{From {\it Spitzer} observation.}
\tablefoottext{c}{Derived from the [S II] $\lambda$6731/$\lambda$6717 line ratio, assuming an electron temperature of T$_e$ = 10000 K.}
}
\end{table*}

\section{Distances and nebular age}\label{s6}

The determination of the distances of Galactic PNs is a notoriously difficult problem. Accurate distances are needed to derive ionized,
molecular, and dust masses of PNs, and the total number of expected Galactic PN population in the Galaxy. One method of distance
determination, which uses extinction-independent parameters has been developed by \citet{Ortiz11}. Making use of the mid-infrared flux and
angular sizes, distances to individual PN can be determined statistically based on a statistical sample correlation coefficients.
This method is particularly appropriate for PNs on the Galactic plane which, suffer from substantial extinction.

The statistical distance as a function of infrared-radio correlation coefficients is given by eq. 4 of \citet{Ortiz11}:

\begin{equation}
\log D_{\rm \scriptscriptstyle MIR}= a_{\rm \scriptscriptstyle 8.3} \log F_{\rm \scriptscriptstyle 8.3} 
+ b_{\rm \scriptscriptstyle 8.3 } \log \theta_{\rm \scriptscriptstyle 8.3}  + c_{\rm \scriptscriptstyle 8.3},
\end{equation}

where D$_{\rm \scriptscriptstyle MIR}$ is the distance of the nebula, F$_{\rm \scriptscriptstyle 8.3}$ is the infrared flux density 
measured from the MSX 8.3 $\mu$m band, $\theta_{\rm \scriptscriptstyle 8.3}$ is the correlated nebular radius obtained from the 
MSX 8.3 $\mu$m image, and $a_{\rm \scriptscriptstyle 8.3}$, $b_{\rm \scriptscriptstyle 8.3}$, and $c_{\rm \scriptscriptstyle 8.3}$ are 
the coefficients listed in the Table \ref{tab1} of \citet{Ortiz11}. According to the measurements of GLIMPSE PNs \citep{Zhang09}, we have
F$_{\rm \scriptscriptstyle 8.3}$ = F$_{\rm \scriptscriptstyle 8.0}$/0.9, where F$_{\rm \scriptscriptstyle 8.0}$ is the {\it IRAC} 
8.0 $\mu$m flux density. Assuming that the nebular radii observed in the MSX 8.3 $\mu$m band ($\theta_{\rm \scriptscriptstyle 8.3}$) are 
the same as that observed in the {\it IRAC} 8.0 $\mu$m channel ($\theta_{\rm \scriptscriptstyle 8.0}$), the formula can be re-written in 
the following form:

\begin{equation}
\log D_{\rm \scriptscriptstyle MIR}= -0.1736 \log F_{\rm \scriptscriptstyle 8.0} -0.3899 \log \theta_{\rm \scriptscriptstyle 8.0}
+ 0.8039.
\end{equation}

The derived distances are given in Table \ref{tab6}. According to \citet{Ortiz11}, the derived distances have error bars $<$ 50$\%$. 

The dynamical ages of these PNs can be estimated by their apparent sizes (from H$\alpha$+[\ion{N}{II}] images), distances, and expansion
velocities. For some bipolar PNs, we adopt the averaged values of their major and minor lengths as their nebular sizes. In the special case 
of PN1, where its edge is not clear, we used the length between extended structures as the adopted size. The distances of the PNs are derived
from the above calculations. An average expansion velocity ($V_{exp}$) of 22.5 km s$^{-1}$ from [\ion{N}{II}] is adopted
\citep{Weinberger89}. The nebular age of each object is then calculated from its physical radius divided by the expansion velocity. 
We estimated the errors of nebular ages based on the statistical errors of the distances (50$\%$, Ortiz et al. 2011). Because the exact 
expansion velocity of each PN is unknown, the derived ages should be taken with some cautions. The nebular expansion velocity is mainly 
affected by the morphological class of a PN \citep{Corradi95}, the effective temperature of the central star ($T_{eff}$), and the mass-loss 
rate of central source (\.{M}) (e.g., Pauldrach et al. 1988). Bipolar PNs are found to have larger nebular expansion velocities than other 
morphological types (elliptical, irregular, point-symmetric, and stellar PNs). Furthermore, the expansion velocity strongly increases with 
$T_{eff}$ and mildly decreases with \.{M}. 

A summary of the properties of these new PNs is also presented in Table \ref{tab6}. We find that most of the PNs are relatively old with an 
averaged age of $\sim$ 8900$\pm$4500 years (compared to the median age of young PNs of 2470 years; Sahai et al. 2011), which is consistent 
with our previous results based on the S2N2 emission ratio diagnostic diagram. For bipolar PNs, the expansion velocities detected in the 
lobe parts are usually faster than in the central regions. Thus, it seems that the derived nebular ages of these PNs are likely to be 
upper limits.

\section{Conclusions}\label{s7}

Recent H$\alpha$ surveys with high spatial resolutions and sensitivities have provided us a powerful tools to study PNs obscured in the
Galactic plane. More than one thousand new PNs and PN candidates have been discovered in the MASH images and IPHAS photometric catalogues
\citep{Mampaso06, Parker06, Miszalski08, Viironen09b, Corradi12}, containing examples of young, compact PNs and old, extended, 
faint PNs. The nature of some of them is still unknown and requires spectroscopy. Here, we present an optical and infrared study of a 
sample of four new PN candidates discovered in the IPHAS and DSH catalogues. We have confirmed the PN status of these objects by the optical 
narrow-band images and mid-resolution spectra. From the locations of the new PNs in the S2N2 emission-line-ratio diagnostic diagram, 
we find that the nebulae are low excitation PNs with high dynamic ages. 

Three new PNs in our sample (PN1, PN2, and PN3) are found to have obvious infrared counterparts. Some of them (PN2 and PN3) display 
prominent infrared emissions in the central regions, suggesting that most dust masses of these objects are located in the central parts. 
The infrared-to-radio flux ratios of these three PNs (PN1, PN2, and PN3) are generally lower than those of other PNs in the GLIMPSE II and 
GLIMPSE 3D fields, suggesting that they are more evolved. We constructed the SED of PN3 by collecting the photometries from a variety of 
astronomical data sets, which cover the wavelength range from visible to far-infrared. The observed infrared fluxes are successfully fitted 
by two dust components with temperatures of 231 K and 60 K, respectively. This PN might have a dusty envelope, which heavily obscures its 
optical counterpart. We estimated the distances and ages of these new PNs. They are relatively old with an averaged age of 
$\sim$ 8900$\pm$4500 years, which is consistent with the results obtained from the S2N2 diagnostic diagram and the infrared-to-radio ratios.

\begin{acknowledgements}

We are grateful to the anonymous referee. His/Her comments helped us to improve this paper.
Part of the data presented in this paper were obtained from the Wide-field Infrared Survey Explorer, which is a joint project of the
University of California, Los Angeles, and the Jet Propulsion Laboratory/California Institute of Technology, funded by the National
Aeronautics and Space Administration; and the {\it Spitzer Space Telescope}, which is operated by the Jet Propulsion Laboratory,
California Institute of Technology under a contract with NASA. This work was partially supported by the Open Project Program of the
Key Laboratory of Optical Astronomy, National Astronomical Observatories, Chinese Academy of Sciences. Financial supports for this work
were provided by the Research Grants Council of the Hong Kong Special Administrative Region, China (project no. HKU 7062/13P and 7073/11P) 
and Small Project Funding of the Hong Kong University (project no. 201007176028).

\end{acknowledgements}

\bibliographystyle{aa}
\bibliography{IPHAS.final}

\begin{thebibliography}{}

\bibitem[Acker et al.(1992)]{Acker92}
Acker, A., Marcout, J., Ochsenbein, F., et al. 1992, Strasbourg-ESO Catalogue of Galactic Planetary Nebulae, Part I and II
(Garching: European Southern Observatory)

\bibitem[Anderson et al.(2012)]{Anderson12}
Anderson, L. D., Zavagno, A., Barlow, M. J., et al. 2012, \aap, 537, 1

\bibitem[Benjamin et al.(2003)]{Benjamin03}
Benjamin, R. A., Churchwell, E., Babler, B. L., et al. 2003, \pasp, 115, 953

\bibitem[Cant\'{o}(1981)]{Canto81}
Cant\'{o}, J. 1981, in Kahn F. D.. ed., Investigating the Universe, Reidel, Dordrecht, p. 95

\bibitem[Churchwell et al.(2009)]{Churchwell09}
Churchwell, E., Babler, B. L., Meade, M. R., et al. 2009, \pasp, 121, 213

\bibitem[Cohen et al.(2005)]{Cohen05}
Cohen, M., Green, A. J., Roberts, M. S. E., et al. 2005, \apj, 627, 446

\bibitem[Cohen et al.(2007)]{Cohen07}
Cohen, M., Parker, Q. A., Green, A. J., et al. 2007, \apj, 669, 343

\bibitem[Cohen et al.(2011)]{Cohen11}
Cohen, M., Parker, Q. A., Green, A. J., et al. 2011, \mnras, 413, 514

\bibitem[Condon \& Kaplan(1998)]{Condon98}
Condon, J. J., \& Kaplan, D. L. 1998, \apjs, 117, 361

\bibitem[Corradi \& Sabin(2012)]{Corradi12}
Corradi, R. L. M., \& Sabin, L. 2012, IAUS, 283, 17 

\bibitem[Corradi \& Schwarz(1995)]{Corradi95}
Corradi, R. L. M., \& Schwarz, H. E. 1995, \aap, 293, 871

\bibitem[Cutri et al.(2003)]{Cutri03}
Cutri, R. M., Skrutskie, M. F., van Dyk, S., et al. 2003, VizieR Online Data Catalog, II/246

\bibitem[Dopita \& Sutherland(2000)]{Dopita00}
Dopita, M. A., \& Sutherland, R. S. 2000, \apj, 539, 742

\bibitem[Draine \& Lee(1984)]{Draine84}
Draine, B. T., \& Lee, H. M. 1984, \apj, 285, 89

\bibitem[Drew et al.(2005)]{Drew05}
Drew, J. E., Greimel, R., \& Irwin, M. J., et al. 2005, \mnras, 362, 753

\bibitem[Fazio et al.(2004)]{Fazio04}
Fazio, G. G., Hora, J. L., Allen, L. E., et al. 2004, \apjs, 154, 10

\bibitem[Filippenko(1982)]{Filippenko82}
Filippenko, A. V. 1982, \pasp, 94, 715

\bibitem[Helou \& Walker(1988)]{Helou88}
Helou, G., \& Walker, D. W. 1988, Infrared Astronomical Satellite (IRAS) Catalogs and Atlases, Vol. 7: The Small Scale Structure 
Catalog (NASA RP-1190; Washington: GPO)

\bibitem[Hsia et al.(2010)]{Hsia10}
Hsia, C.-H., Kwok, S., Zhang, Y., et al. 2010, \apj, 725, 173

\bibitem[Howarth(1983)]{Howarth83}
Howarth, I. D. 1983, \mnras, 203, 301 

\bibitem[Hummer \& Storey(1987)]{Hummer87}
Hummer, D. G., \& Storey, P. J. 1987, \mnras, 224, 801

\bibitem[Ishihara et al.(2010)]{Ishihara10}
Ishihara, D., Onaka, T., Kataza, H., et al. 2010, \aap, 514, A1

\bibitem[Jacoby et al.(2010)]{Jacoby10}
Jacoby, G. H., Kronberger, M., Patchick, D., et al. 2010, \pasa, 27, 156

\bibitem[Kohoutek(2001)]{Kohoutek01}
Kohoutek, L. 2001, \aap, 378, 843

\bibitem[Kronberger et al.(2006)]{Kronberger06}
Kronberger, M., Teutsch, P., Alessi, B., et al. 2006, \aap, 447, 921


\bibitem[Kwok(1989)]{Kwok89}
Kwok, S. 1989, in IAU Symp. 131, Planetary Nebulae, ed. S. Torres-Peimbert (Kluwer: Reidel), 401

\bibitem[Kwok(2007)]{Kwok07}
Kwok, S. 2007, Physics and Chemistry of the Interstellar Medium (Sausalito, CA:Univ. Science Books)

\bibitem[Kwok et al.(2008)]{Kwok08}
Kwok, S., Zhang, Y., Koning, N., et al. 2008, \apjs, 174, 426


\bibitem[Luna \& Costa(2005)]{Luna05}
Luna, G. J. M., \& Costa, R. D. D. 2005, \aap, 435, 1087

\bibitem[Mampaso et al.(2006)]{Mampaso06}
Mampaso, A., Corradi, R. L. M., Viironen, K. 2006, \aap, 458, 203

\bibitem[Mathis et al.(1977)]{Mathis77}
Mathis, J. S., Rumpl, W., \& Nordsieck, K. H. 1977, \apj, 217, 425

\bibitem[Miszalski et al.(2008)]{Miszalski08}
Miszalski, B., Parker, Q. A., Acker, A., et al. 2008, \mnras, 384, 525

\bibitem[Murakami et al.(2007)]{Murakami07}
Murakami, H., Baba, H., Barthel, P., et al. 2007, \pasj, 59, 369 

\bibitem[Neugebauer et al.(1984)]{Neugebauer84}
Neugebauer, G., Habing, H. J., van Duinen, R., et al. 1984, \apj, 278, L1

\bibitem[Ortiz et al.(2011)]{Ortiz11}
Ortiz, R., Copetti, M. V. F., \& Lorenz-Martins, S. 2011, \mnras, 418, 2004

\bibitem[Parker et al.(2006)]{Parker06}
Parker, Q. A., Acker, A., Frew, D. J., et al. 2006, \mnras, 373, 79

\bibitem[Pauldrach et al.(1988)]{Pauldrach88}
Pauldrach, A., Puls, J., Kudritzki, R.-P., et al. 1988, \aap, 207, 123

\bibitem[Pottasch et al.(1984)]{Pottasch84}
Pottasch, S. R., Baud, B., Beintema, D., et al. 1984, \aap, 138, 10


\bibitem[Reach et al.(2005)]{Reach05}
Reach, W. T., Megeath, S. T., Cohen, M., et al. 2005, \pasp, 117, 978

\bibitem[Raga et al.(1996)]{Raga96}
Raga, A. C., B\"{o}hm, K. -H., \& Cant\'{o}, J. 1996, \rmxaa, 32, 161

\bibitem[Rieke et al.(2004)]{Rieke04}
Rieke, G. H., Young, E. T., Engelbracht, C. W., et al. 2004, \apjs, 154, 25

\bibitem[Riesgo \& L\'{o}pez(2006)]{Riesgo06}
Riesgo, H., \& L\'{o}pez, J. A. 2006, \rmxaa, 42, 47

\bibitem[Sabbadin et al.(1977)]{Sabbadin77}
Sabbadin, F., Minello, S., \& Bianchini, A. 1977, \aap,  60, 147

\bibitem[Sabin et al.(2010)]{Sabin10}
Sabin, L., Zijlstra, A. A., Wareing, C., et al. 2010, \pasa, 27, 166

\bibitem[Sabin et al.(2011)]{Sabin11}
Sabin, L., Corradi, R. L. M., \& Mampaso, A. 2011, RMxAC, 40, 183

\bibitem[Sahai et al.(2011)]{Sahai11}
Sahai, R., Morris, M. R., \& Villar, G. G. 2011, \aj, 141, 134

\bibitem[Skrutskie et al.(2006)]{Skrutskie06}
Skrutskie, M. F., Cutri, R. M., Stiening, R., et al. 2006, \aj, 131, 1163 

\bibitem[Stasi\'{n}ska \& Szczerba(1999)]{Stasinska99}
Stasi\'{n}ska, G., \& Szczerba, R. 1999, \aap, 352, 297

\bibitem[Storey \& Zeippen(2000)]{Storey00}
Storey, P. J. \& Zeippen, C. J. 2000, \mnras, 312, 813

\bibitem[Viironen et al.(2007)]{Viironen07}
Viironen, K., Delgado-Inglada, G., Mampaso, A., et al. 2007, \mnras, 381, 1719

\bibitem[Viironen et al.(2009a)]{Viironen09a}
Viironen, K., Greimel, R., Corradi, R. L. M., et al. 2009a, \aap, 504, 291

\bibitem[Viironen et al.(2009b)]{Viironen09b}
Viironen, K., Mampaso, A., Corradi, R. L. M., et al. 2009b, \aap, 502, 113

\bibitem[Wright et al.(2010)]{Wright10}
Wright, E. L., Eisenhardt, P. R. M., Mainzer, A. K., et al. 2010, \aj, 140, 1868

\bibitem[Weinberger(1989)]{Weinberger89}
Weinberger, R. 1989, \aaps, 78, 301

\bibitem[Zhang \& Kwok(1991)]{Zhang91}
Zhang, C. Y., \& Kwok, S. 1991, \aap, 250, 179

\bibitem[Zhang \& Kwok(2009)]{Zhang09}
Zhang, Y., \& Kwok, S. 2009, \apj, 706, 252

\bibitem[Zhang et al.(2012)]{Zhang12}
Zhang, Y., Hsia, C. -H., \& Kwok, S. 2012, \apj, 745, 59

\end{thebibliography}

\end{document}